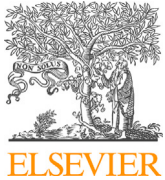
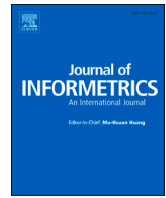

Research Paper

# Gini-stable Lorenz curves and their relation to the generalised Pareto distribution

Lucio Bertoli-Barsotti [a], Marek Gagolewski [b,c,e,*], Grzegorz Siudem [d], Barbara Żogała-Siudem [e]

[a] *University of Bergamo, Department of Economics, Italy*
[b] *Deakin University, Data to Intelligence Research Centre, School of IT, Geelong, VIC 3220, Australia*
[c] *Warsaw University of Technology, Faculty of Mathematics and Information Science, ul. Koszykowa 75, 00-662 Warsaw, Poland*
[d] *Warsaw University of Technology, Faculty of Physics, ul. Koszykowa 75, 00-662 Warsaw, Poland*
[e] *Systems Research Institute, Polish Academy of Sciences, ul. Newelska 6, 01-447 Warsaw, Poland*

A B S T R A C T

We introduce an iterative discrete information production process where we can extend ordered normalised vectors by new elements based on a simple affine transformation, while preserving the predefined level of inequality, G, as measured by the Gini index.

Then, we derive the family of empirical Lorenz curves of the corresponding vectors and prove that it is stochastically ordered with respect to both the sample size and G which plays the role of the uncertainty parameter. We prove that asymptotically, we obtain all, and only, Lorenz curves generated by a new, intuitive parametrisation of the finite-mean Pickands' Generalised Pareto Distribution (GPD) that unifies three other families, namely: the Pareto Type II, exponential, and scaled beta distributions. The family is not only totally ordered with respect to the parameter G, but also, thanks to our derivations, has a nice underlying interpretation. Our result may thus shed a new light on the genesis of this family of distributions.

Our model fits bibliometric, informetric, socioeconomic, and environmental data reasonably well. It is quite user-friendly for it only depends on the sample size and its Gini index.

## 1. Introduction

The well-known information production process (IPP; see Egghe, 1990 and Egghe, 2005b, p. 8) is a mechanism in which "sources" (e.g., authors, universities) produce a series of "items" (e.g., articles, graduates) of different quality. In the "source-item" formulation, the variable of interest is typically non-negative and integer-valued for both the rank-size and rank-frequency modelling (Bertoli-Barsotti & Lando, 2019, Egghe & Waltman, 2011). In spite of this, in the informetric literature, continuous approximations are typically used because it facilitates the underlying calculations (Burrell, 2005, Egghe, 2005a). However, in this article, we consider a natural framework for empirical data stemming from a variety of applications in informetrics that entails the *discrete* setting.

* Corresponding author at: Deakin University, Data to Intelligence Research Centre, School of IT, Geelong, VIC 3220, Australia.
  *E-mail addresses:* lucio.bertoli-barsotti@unibg.it (L. Bertoli-Barsotti), m.gagolewski@deakin.edu.au (M. Gagolewski), grzegorz.siudem@pw.edu.pl (G. Siudem), zogala@ibspan.waw.pl (B. Żogała-Siudem).
  *URLs:* https://www.gagolewski.com (M. Gagolewski), http://if.pw.edu.pl/ (G. Siudem).

Lorenz curves are important instruments in operational research (e.g., decision making and resource allocation; Argyris et al., 2022, Bu, 2022), economics (Aaberge, 2001, Gajdos, 2004, Magdalou, 2021), bibliometrics, and informetrics (Burrell, 1992, 2006, 2007, Egghe, 2005c, Egghe & Rousseau, 2023, Prathap, 2023, Prathap & Rousseau, 2023, Rousseau et al., 2023, Sarabia, 2008); see Section 2 for a review of the terminology. In this paper, we introduce a new parametric model that consists of Lorenz curves $L^{(N,G)}$ of a vector of $N$ positive numbers (without loss of generality, we will be studying $N$-dimensional probability vectors). By construction, $L^{(N,G)}$ will be a piecewise linear function depending on a free parameter $G$ with a straightforward interpretation. Namely, $G$ represents the most direct normalised measure of inequality of the components of the vector: the Gini index. Our model gives an explicit expression for the Lorenz curve for all the values of the Gini index in the range $(0,1)$, which makes it very flexible. Moreover, we prove that $L^{(N,G)}$ defines a family of Lorenz curves which are Lorenz-ordered with respect to both parameters, $N$ and $G$. Thus, if $G_1 < G_2$, then $L^{(N,G_1)}(u) \geq L^{(N,G_2)}(u)$, and if $N_1 < N_2$, then $L^{(N_1,G)}(u) \geq L^{(N_2,G)}(u)$ for all $u$.

Apart from presenting a new model of Lorenz curves $L^{(N,G)}$ of a vector of finite dimension $N$, we will be interested in studying its characteristics as the dimension diverges to infinity. In the limit as $N \to \infty$, the model becomes governed by only one parameter: the inequality index $G$. This limit defines a new parametric family of Lorenz curves $L^{(\infty,G)}$, ordered by $G$. Thus, if $G_1 < G_2$, then $L^{(\infty,G_1)}(u) \geq L^{(\infty,G_2)}(u)$ for all $u$. Interestingly, the family admits a nice characterisation result. Namely, we prove that it contains all, and only, Lorenz curves generated by a Generalised Pareto Distribution (for which the mean exists).

In order to define the model $L^{(N,G)}$, as an ancillary result, in Section 3, we introduce the notion of the *Gini-stable* process, i.e., a sequence of ordered and normalised vectors with a fixed (stable) value $G$ of the Gini index. In particular, we mention that $G$ is an "uncertainty parameter" of the Gini-stable process. In Section 4, we present the results concerning the limiting family of Lorenz curves. Then, we prove the aforementioned ordering properties of our curve set. To strengthen our theoretical derivations, we present a few case studies, including the cases of socioeconomic, bibliometric, informetric, and environmental data. We note that, more often than not, our model fits empirical data reasonably well.

## 2. Preliminaries

### 2.1. Lorenz curve and Lorenz order

Let $\wp_+^{(N)}$ be the set of $N$-dimensional ordered[1] and normalised vectors like $\boldsymbol{p} = (p_1, p_2, \ldots, p_N)$, i.e., with

$$p_1 \geq p_2 \geq \cdots \geq p_N > 0 \text{ and } \sum_{i=1}^{N} p_i = 1.$$

Furthermore, given $\boldsymbol{p} \in \wp_+^{(N)}$, let $\Sigma_i^p = \sum_{j=1}^{i} p_j$ denote its $i$-th cumulative sum, with, for the brevity of the anticipated formulae, $\Sigma_0^p = 0$.

Assume we are given a set of numbers $\{p_1, p_2, \ldots, p_N\}$, which are featured as the components of some $\boldsymbol{p} \in \wp_+^{(N)}$. Then, the corresponding *Lorenz curve* is the function $L^p(u)$, $0 \leq u \leq 1$, obtained by linear interpolation of the points $\left(\frac{i}{N}, L^p\left(\frac{i}{N}\right)\right)$, where for $i = 0, 1, \ldots, N$, $L^p\left(\frac{i}{N}\right) = 1 - \Sigma_{N-i}^p$, i.e., the sum of the first $i$ smallest elements; see (Bertoli-Barsotti & Lando, 2019, Eq. (3)) and Fig. 1 for an example.

As well known, the *Lorenz order* is defined in terms of nested Lorenz curves as follows. For any $\boldsymbol{p}, \boldsymbol{q} \in \wp_+^{(N)}$, $\boldsymbol{p} \leq_L \boldsymbol{q}$ if and only if $L^p(u) \geq L^q(u)$ for all $0 \leq u \leq 1$.

### 2.2. Gini index of a probability vector

Let us recall some basic facts related to the sample Gini (1912) index, which for $\boldsymbol{p} \in \wp_+^{(N)}$ is given by:

$$\mathcal{G}(\boldsymbol{p}) = \frac{1}{N-1} \sum_{i=1}^{N-1} \sum_{j=i+1}^{N} |p_i - p_j|.$$

The index is normalised so that $\mathcal{G}(1, 0, \ldots, 0) = 1$ and $\mathcal{G}(1/N, 1/N, \ldots, 1/N) = 0$.

Following (Pyatt, 1976), see also (Tumen, 2011), in the context of wealth distribution, where a given $p_i$ represents the proportion of the overall wealth that is held by the $i$-th richest individual, we may interpret the quantity $\max\{0, p_j - p_i\}$ as the "gain" of the individual $i$ when given the option of being someone else, $j \neq i$. Then, we can consider $T_i = \frac{\sum_{j=1, j \neq i}^{N} \max\{0, p_j - p_i\}}{(N-1)}$ as the expected gain of individual $i$. We find that the average expected gain $\frac{1}{N} \sum_{i=1}^{N} T_i$ is equal to $\frac{1}{N} \frac{\sum_{i=1}^{N-1} \sum_{j=i+1}^{N} |p_i - p_j|}{(N-1)}$. Dividing the average expected gain by the mean, $N^{-1}$ (as our vectors sum to 1), we get that the average relative expected gain is $N\frac{\sum_{i=1}^{N} T_i}{N} = \sum_{i=1}^{N} T_i = \mathcal{G}(p)$. Hence, the Gini index $\mathcal{G}(p)$ can be seen as *the sum of the average expected gains* $T_i$, $i = 1, 2, \ldots, N$.

---

[1] We follow the convention from, amongst others, (Egghe & Waltman, 2011) where it is assumed that the vectors are nonincreasingly ordered, i.e., the highest-valued item is listed first.

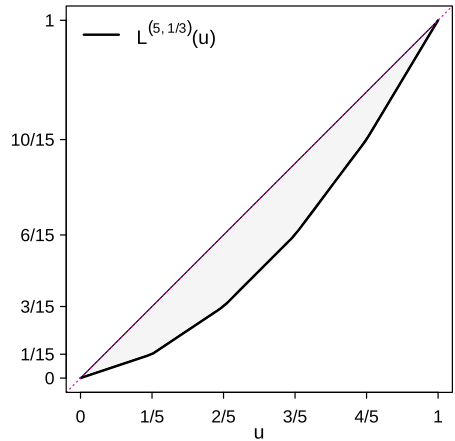

**Fig. 1.** The Lorenz curve $L^{(5,1/3)}$ of an example probability vector of length $N = 5$ generated from our model (Section 3), $p^{(5,1/3)} = (5/15, 4/15, 3/15, 2/15, 1/15)$. The Gini index of this vector, $\mathcal{G}(p^{(5,1/3)}) = 1/3$, is equal to $2N/(N-1)$ times the area between the identity (45-degree) line and the Lorenz curve.

**Example 1.** If $p = (5/15, 4/15, 3/15, 2/15, 1/15)$ as in Fig. 1, we find the average relative expected gains: $T_1 = 0$, $T_2 = (1/15)/4 = 1/60$, $T_3 = (2/15 + 1/15)/4 = 3/60$, $T_4 = (3/15 + 2/15 + 1/15)/4 = 6/60$, $T_5 = (4/15 + 3/15 + 2/15 + 1/15)/4 = 10/60$. Then, $\mathcal{G}(p) = 1/60 + 3/60 + 6/60 + 10/60 = 20/60 = 1/3$.

A useful alternative formula for the Gini index is:

$$\mathcal{G}(p) = \frac{2\hat{\mu} - N - 1}{N - 1}, \quad (1)$$

where $\hat{\mu}$ is the mean (expected) rank, that is, $\hat{\mu} = \sum_{i=1}^{N}(N - i + 1)p_i$ (compare Arnold, 2015, p. 183). What is more, the Gini index is closely related to the Lorenz curve through the following formula with a clear geometrical meaning:

$$\mathcal{G}(p) = 2\frac{N}{N-1}A(p),$$

where $A(p) = \int_0^1 (u - L^p(u)) \, du$ is the area between the Lorenz curve of the given vector and the graph of the identity line; compare Fig. 1.

### 2.3. Affine inequality-attenuating transformations

In the context of the Lorenz order, a function $h : \mathbb{R}^+ \to \mathbb{R}^+$ is said to be *inequality-attenuating*, whenever $h(p) \leq_L p$, where the componentwise extension of $h$ is $h(p) = (h(p_1), h(p_2), \ldots, h(p_N))$; see Arnold (1991).

Let us consider the affine function $h : \mathbb{R}^+ \to \mathbb{R}^+$ of the type:

$$h(p_i) = a + bp_i, \qquad i = 1, 2, \ldots, N,$$

where $0 < b < 1$, and $a = \frac{1-b}{N}$, in order to have $\sum_{i=1}^{N} h(p_i) = 1$. This function defines a reduce-and-distribute mapping from $\wp_+^{(N)}$ to $\wp_+^{(N)}$: it proportionally *reduces* each component by a factor of $b$ and *distributes* the same amount $a$ to every element. It is easy to see that this transformation is inequality-attenuating in that $h(p) \leq_L p$ for every $p$ (for a general characterisation result on inequality-attenuating functions, refer to Marshall et al., 2011, p. 727 and Arnold & Sarabia, 2018, p. 36–37). In particular, it decreases the expected gain $T_i$ of each individual $i$ by a constant factor $bT_i$. Indeed, the Gini index decreases:

$$G = \mathcal{G}(p_1, p_2, \ldots, p_N) > \mathcal{G}(a + bp_1, a + bp_2, \ldots, a + bp_N) = bG.$$

## 3. The Gini-stable process and distribution

### 3.1. Gini-stable transformations

Let us now introduce a slightly different type of mapping: the one in which we extend the dimension of the vector, from $N$ to $N + 1$:

$$(p_1, p_2, \ldots, p_N) \xrightarrow{h} (a + bp_1, a + bp_2, \ldots, a + bp_N, a), \qquad (2)$$

where $a = \frac{1-b}{N+1}$, in order to have $a + \sum_{i=1}^{N}(a + bp_i) = 1$.

The mapping given by Eq. (2) is from $\wp_+^{(N)}$ to $\wp_+^{(N+1)}$, and it may no longer have an inequality-attenuating effect; for example, as $b$ tends to 1, the Lorenz curves of $\boldsymbol{p}$ and $h(\boldsymbol{p})$ are nested.

Instead, we can identify the constant $b = b_N$ (which in general depends on the dimension of the vector) uniquely determined by the constraint:

$$\mathcal{G}(\boldsymbol{p}) = G = \mathcal{G}(h(\boldsymbol{p})), \qquad \boldsymbol{p} \in \wp_+^{(N)}, \ h(\boldsymbol{p}) \in \wp_+^{(N+1)},$$

that balances the natural inequality-attenuating effect of $h$, and the inequality-increasing effect due to the increase in the vector size. In other words, our purpose is to determine the affine function $h$ which *keeps the value $G$ of the Gini index unchanged when a new component is added to the vector*.

The general solution of this problem is a process formed by a sequence of probability $N$-vectors, $\boldsymbol{p}^{(N,G)}$, $N = 2, 3, \ldots$, that guarantees a "stable" (the same, fixed, invariant) value of $G$, i.e., $\mathcal{G}(\boldsymbol{p}^{(N,G)}) = G$, for every fixed $G \in (0, 1)$.

In an economic context, the task at hand can be stated as the problem of finding the conditions under which the sum of the average expected gains in a population of $N$ individuals remains unchanged when the size of the population increases by one, from $N$ to $N + 1$. This issue is addressed by an iterative approach, that we will be referring to as the *Gini-stable* process.

Note that in the base case of $N = 2$, the probability vector is *uniquely* determined by $G$. Namely, by virtue of Eq. (1), we find $\mathcal{G}(\boldsymbol{p}^{(2,G)}) = 3 - 2\hat{\mu}$, where $\hat{\mu} = p_1^{(2,G)} + 2p_2^{(2,G)} = p_1^{(2,G)} + 2(1 - p_1^{(2,G)}) = 2 - p_1^{(2,G)}$. Then, $\boldsymbol{p}^{(2,G)}$ can be expressed as a function of $G$, $0 < G < 1$, as $\boldsymbol{p}^{(2,G)} = \left(\frac{1+G}{2}, \frac{1-G}{2}\right)$.

When $N \geq 3$, there are infinitely many probability $N$-vectors that determine the same value $G$ of the Gini index, but under the affine constraint given by the condition (2), the solution becomes unique. This is because the condition $G = \mathcal{G}(\boldsymbol{p}^{(N+1,G)})$ is equivalent to $S(\boldsymbol{p}^{(N+1,G)}) = 2NG$, where:

$$S(\boldsymbol{p}^{(N+1,G)}) = \sum_{i=1}^{N}\sum_{j=1}^{N} \left|p_i^{(N+1,G)} - p_j^{(N+1,G)}\right| + 2\sum_{i=1}^{N}\left(p_i^{(N+1,G)} - p_{N+1}^{(N+1,G)}\right)$$

$$= S(\boldsymbol{p}^{(N,G)})b_N + 2\sum_{i=1}^{N} b_N p_i^{(N,G)} = 2(N-1)Gb_N + 2b_N.$$

Then, we find the equivalent equation $2NG = 2(N-1)Gb_N + 2b_N$, that leads to the solution:

$$b_N = \frac{NG}{NG + 1 - G}. \qquad (3)$$

In conclusion, we introduced a sequence of vectors with components defined in terms of the following iterative equations:

$$p_i^{(N+1,G)} = \frac{1-G}{(N+1)(NG+1-G)} + \frac{NG}{NG+1-G} p_i^{(N,G)}, \qquad \text{for } i = 1, 2, \ldots, N, \qquad (4)$$

$$p_{N+1}^{(N+1,G)} = \frac{1-G}{(N+1)(NG+1-G)}.$$

**Example 2.** If $G = 1/3$, we have: $\boldsymbol{p}^{(2,G)} = \left(\frac{1+G}{2}, \frac{1-G}{2}\right) = \left(\frac{2}{3}, \frac{1}{3}\right)$. Then, for $N + 1 = 3$, we find:

$$a_2 = \frac{1-G}{(N+1)(NG+1-G)} = \frac{2/3}{3(2/3+1-1/3)} = \frac{1}{6},$$

$$b_2 = \frac{NG}{NG+1-G} = \frac{2/3}{2/3+1-1/3} = \frac{3}{6}.$$

Hence, $\boldsymbol{p}^{(3,1/3)} = (1/6 + 3/6 \cdot 2/3, 1/6 + 3/6 \cdot 1/3, 1/6) = (3/6, 2/6, 1/6)$.
Subsequently, for $N + 1 = 4$ we find $a_3 = 1/10$, $b_3 = 6/10$, and $\boldsymbol{p}^{(4,1/3)} = (4/10, 3/10, 2/10, 1/10)$.
Moreover, $N + 1 = 5$ yields $a_4 = 1/15$, $b_4 = 10/15$, and $\boldsymbol{p}^{(5,1/3)} = (5/15, 4/15, 3/15, 2/15, 1/15)$, and so forth.
All the above vectors have the same Gini index, $G = 1/3$.

It turns out (see Appendix A.1 for the derivation) that our iterative model given by Eq. (4) enjoys the analytic formula:

$$p_i^{(N,G)} = \begin{cases} \frac{1-G}{2G-1} \frac{1}{N} \left( \frac{\Gamma(N+1)\Gamma(i-2+1/G)}{\Gamma(N-1+1/G)\Gamma(i)} - 1 \right), & G \neq \frac{1}{2}, \\ \frac{1}{N}(H_N - H_{i-1}) = \frac{1}{N}\sum_{j=i}^{N} \frac{1}{j}, & G = \frac{1}{2}, \end{cases} \qquad (5)$$

where $i = 1, 2, \ldots, N$, and $H_N = \sum_{j=1}^{N} \frac{1}{j}$ is the $N$-th harmonic number, $H_0 = 0$. Note that the case $G = 1/2$ must have been considered separately as otherwise it would lead to a division by 0.

We will call the vector $\boldsymbol{p}^{(N,G)}$ a *Gini-stable distribution* with parameters $N$ and $G$, denoted GSD$(N, G)$. Nevertheless, let us note that our process is completely deterministic. Thence, it should not be interpreted in probabilistic terms (without making further assumptions).

Take note of the limit cases: as $G \to 0$, we obtain the discrete uniform distribution with support $1, 2, \ldots, N$ (all components being equal), and as $G \to 1$, we get the degenerate distribution with all probability mass concentrated at the point $i = 1$.

*3.2. Lorenz curves of the vectors generated by the Gini-stable process*

An interesting property of the Gini-stable distribution is that it admits an explicit representation of its cumulative sums. This allows us to obtain the Lorenz curve corresponding to the vector $\boldsymbol{p}^{(N,G)}$. Namely, starting from Eq. (5), for $i = 1, 2, \ldots, N$, after some elementary transformations, we get:

$$\Sigma_i^{(N,G)} = \sum_{j=1}^{i} p_j^{(N,G)} = \begin{cases} \frac{1-G}{2G-1} \left( \frac{G}{1-G} \frac{\Gamma(N)\Gamma(i+1/G-1)}{\Gamma(i)\Gamma(N+1/G-1)} - \frac{i}{N} \right), & G \neq \frac{1}{2}, \\ \frac{i}{N} \left( 1 + H_N - H_i \right), & G = \frac{1}{2}. \end{cases} \quad (6)$$

Based on the above, the Lorenz curves of $\boldsymbol{p}^{(N,G)}$ are defined for $i = 1, 2, \ldots, N$ as follows:

$$\begin{aligned} L^{(N,G)}\left(\frac{i}{N}\right) &= 1 - \Sigma_{N-i}^{(N,G)} \\ &= \begin{cases} 1 - \frac{1-G}{2G-1} \left( \frac{G}{1-G} \frac{\Gamma(N)}{\Gamma(N-1+1/G)} \frac{\Gamma(N-i-1+1/G)}{\Gamma(N-i)} - \frac{N-i}{N} \right), & \text{if } G \neq \frac{1}{2}, \\ 1 - \frac{N-i}{N} \left( 1 + H_N - H_{N-i} \right), & \text{if } G = \frac{1}{2}. \end{cases} \end{aligned} \quad (7)$$

with the values in-between linearly interpolating the above points as described in Section 2.

**Example 3.** Let us consider the Luxembourg Income Study (LIS) dataset as per Bishop et al. (1991, Table 2). It gives empirical Lorenz curves, across the deciles ($N = 10$), for the family incomes in nine different countries, $L^{(\mathrm{AU})}$, $L^{(\mathrm{CA})}$, etc. By computing the corresponding Gini indices $\hat{G}_{\mathrm{AU}}, \hat{G}_{\mathrm{CA}}, \ldots$, we get the predicted models $L^{(N,\hat{G}_{\mathrm{AU}})}$, $L^{(N,\hat{G}_{\mathrm{CA}})}$, etc., using the derived formula. In Fig. 2, we see that our model fits most of the empirical curves very well: in five cases, the maximal root mean squared error (RMSE) is not greater than 0.013, and no RMSE is greater than 0.027.

*3.3. Lorenz ordering for vectors of equal lengths*

Let $\boldsymbol{p}^{(N,G)}$ be a vector generated by the Gini-stable process, i.e., with components given by Eq. (5), and let $L^{(N,G)}$ be its Lorenz curve, as defined by Eq. (7). The following proposition shows that, for any fixed $N$, an increase of the parameter $G$ corresponds to an increase in the Lorenz ordering (see Fig. 3).

**Proposition 1.** *For every fixed $N \geq 2$, if $G_1 < G_2$, then $L^{(N,G_1)}(u) > L^{(N,G_2)}(u)$ for all $u \in (0, 1)$.*

**Proof.** See Appendix A.2. □

This result can be rephrased by saying that $\boldsymbol{p}^{(N,G_1)} \leq_L \boldsymbol{p}^{(N,G_2)}$, whenever $G_1 < G_2$.

**Remark 1.** The condition $1 - L^p \left( 1 - \frac{i}{N} \right) = \Sigma_i^p = \sum_{k=1}^{i} p_k \leq \sum_{k=1}^{i} q_k = \Sigma_i^q = 1 - L^q \left( 1 - \frac{i}{N} \right)$, $i = 0, 1, \ldots, N-1$, $\sum_{i=1}^{N} p_i = \sum_{i=1}^{N} q_i = 1$, is the partial sum definition of the majorisation $\boldsymbol{p} \prec \boldsymbol{q}$ (see Marshall et al., 2011, p. 8). The Lorenz order $\leq_L$ is more general than majorisation, because majorisation does not apply to vectors of unequal lengths. Nevertheless, the Lorenz order and majorisation coincide within the set $\wp_+^N$, in that for every pair $\boldsymbol{p}, \boldsymbol{q} \in \wp_+^N$, $\boldsymbol{p} \prec \boldsymbol{q}$ if and only if $\boldsymbol{p} \leq_L \boldsymbol{q}$. We can interpret the above relation by saying that the vector $\boldsymbol{p}$ is *more random* than $\boldsymbol{q}$ (see Marshall et al., 2011, p. 754).

Hence, Proposition 1 is equivalent to stating that for $G_1 < G_2$, it holds $\boldsymbol{p}^{(N,G_1)} \prec \boldsymbol{p}^{(N,G_2)}$, where $\prec$ is the majorisation ordering (compare Bertoli-Barsotti, 2023).

This implies that $G$ is an "uncertainty parameter" of the Gini-stable distribution GSD$(N, G)$ in the sense of Hickey (1983), in that the Gini-stable distribution GSD$(N, G_1)$ possesses a degree of randomness greater than that of the Gini-stable distribution GSD$(N, G_2)$ (compare Marshall et al., 2011, p. 755).

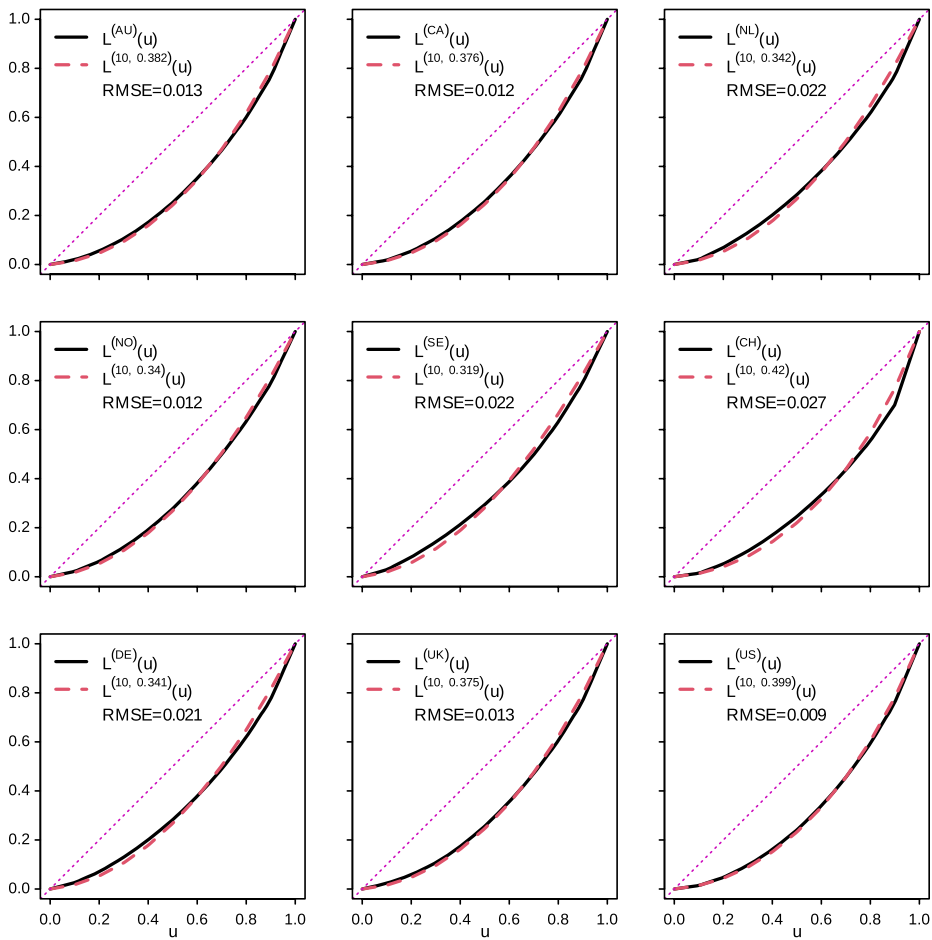

**Fig. 2.** The empirical and fitted curves for family income data from (Bishop et al., 1991, Table 2); see Examples 3 and 5. In most cases, our Gini-stable distribution gives a very good fit.

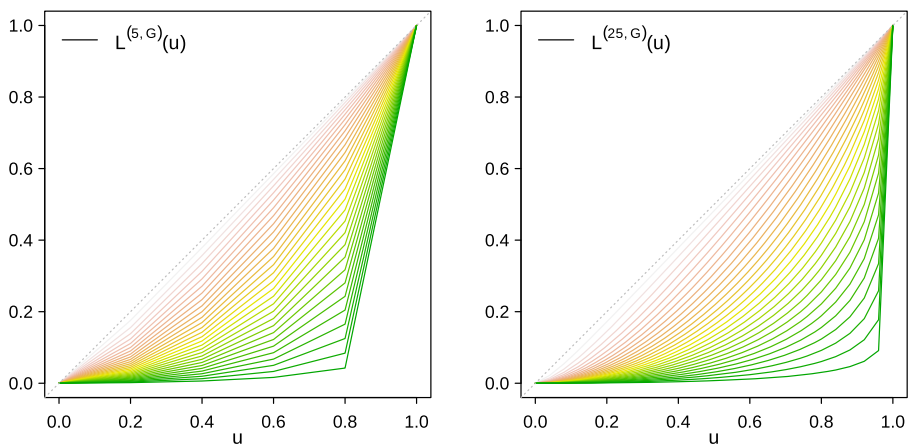

**Fig. 3.** Lorenz curves $L^{(N,G)}$ for $N=5$ (left) and $N=25$ (right) and different values of the parameter $G = 1/30, 2/30, \ldots, 29/30$ (from brown through yellow at $G = 1/2$ to green). The curves are Lorenz-ordered, as per Proposition 1.

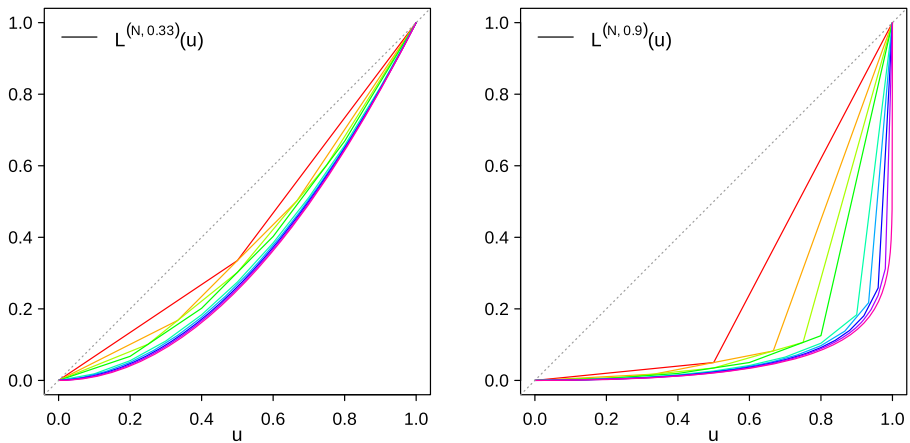

**Fig. 4.** Lorenz curves $L^{(N,G)}$ for $G = 1/3$ (left) and $G = 0.9$ (right) and different values of the parameter $N = 2, 3, 4, 5, 10, 15, 25, 50, \infty$ (from red through cyan to purple). The limiting case $L^{(\infty,G)}$ is given by Eq. (8). The curves converge towards the bottom-right corner: they form a chain with respect to the Lorenz order (Proposition 2).

### 3.4. Lorenz ordering for vectors of unequal lengths

Lorenz order allows to compare inequality in populations of different sizes. In our model, for fixed $G$, an increase of $N$ corresponds to an increase in the Lorenz ordering (see Fig. 4).

**Proposition 2.** *For every fixed $G$, $0 < G < 1$ and $u$, $0 \le u \le 1$, $L^{(N_1,G)}(u) \ge L^{(N_2,G)}(u)$, whenever $N_1 < N_2$.*

**Proof.** See Appendix A.3. □

Proposition 2 can be interpreted by saying that $\boldsymbol{p}^{(N_1,G)} \le_L \boldsymbol{p}^{(N_2,G)}$ whenever $N_1 < N_2$.

## 4. The limiting case as $N \to \infty$

### 4.1. Family of Lorenz curves, $L^{(\infty,G)}$

As $N$ grows indefinitely, the vector $\boldsymbol{p}^{(N,G)}$ becomes an infinite sequence of numbers $(p_1, p_2, \dots)$.

Then, the Lorenz curves take an asymptotic form that depends on only one parameter, $G$. Namely, we can prove (see Appendix A.4) that they can be expressed for $u \in [0,1]$ by:

$$L^{(\infty,G)}(u) = \lim_{N \to \infty} L^{(N,G)}(u) = \begin{cases} u + \dfrac{G}{2G-1} \left(1 - u - (1-u)^{1/G-1}\right), & \text{if } G \ne \tfrac{1}{2}, \\ u + (1-u)\log(1-u), & \text{if } G = \tfrac{1}{2}. \end{cases} \tag{8}$$

See Fig. 4 again for an illustration.

**Remark 2.** We easily find that $\int_0^1 L^{(\infty,G)} \, du = \frac{1-G}{2}$. Thence, $2 \int_0^1 \left(u - L^{(\infty,G)}\right) du = G$, as expected in the continuous case.

As an extension of Propositions 1 and 2, we have the result below.

**Proposition 3.** *The following inequalities hold:*

(i) *If $G_1 < G_2 < 1$, then $L^{(\infty,G_1)}(u) \ge L^{(\infty,G_2)}(u)$ for all $u \in [0,1]$,*
(ii) *For every fixed $G \in (0,1)$ and any $N \ge 2$, $L^{(N,G)}(u) \ge L^{(\infty,G)}(u)$ for all $u \in [0,1]$.*

**Proof.** The result *(i)* follows from the observation that $\frac{\partial L^{(\infty,G)}(u)}{\partial G} < 0$. The result *(ii)* follows from Proposition 2. □

Fig. 5 depicts the limiting Lorenz curves for different values of the Gini index, $G$. Quite remarkably, these curves cover the entire family of distributions known as Pickands' Generalised Pareto Distribution, as we shall show next.

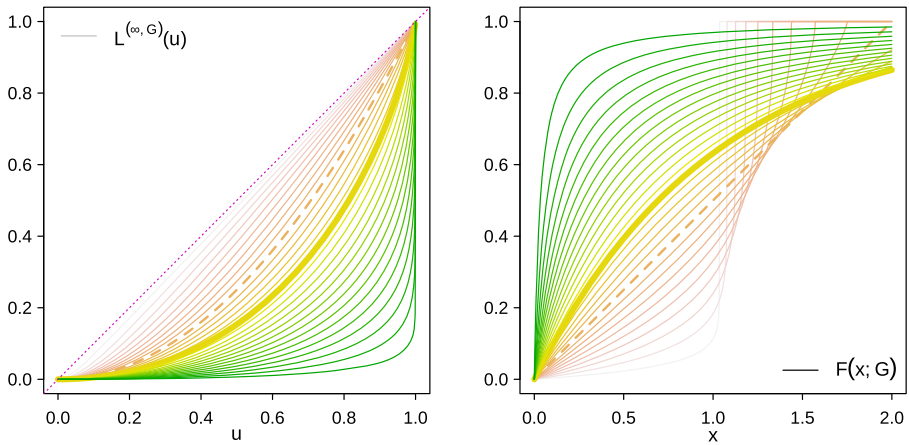

**Fig. 5.** Left: Lorenz curves $L^{(\infty,G)}$, for $G = 1/30, 2/30, \ldots, 29/30$ (from brown to green). Right: The CDF of the generalised Pareto distributions given by Eq. (10) utilising our new parametrisation, $F(x; G)$ (with expectation 1, i.e., $m = 1$). The thicker yellow curve represents the case $G = 1/2$, corresponding to the exponential distribution. The brown-yellow curves correspond to scaled Beta distributions (with the dashed line denoting the uniform distribution), whilst the yellow-green ones correspond to Pareto Type II distributions.

**Example 4.** Let us consider a subset of data analysed by Clauset et al. (2009) that includes the following nine datasets available for download from the author's website[2]: *blackouts* (number of customers affected by electrical blackouts in the US between 1984 and 2002), *quakes* (maximal amplitudes of Californian earthquakes between 1910 and 1992), *words* (number of occurrences of unique words in H. Melville's "Moby Dick"), *cities* (populations of US cities in the 2000 US Census), *flares* (peak $\gamma$ ray intensity of solar flares between 1980 and 1989), *fires* (sizes of wildfires on the US federal land between 1986 and 1996, in acres), *terrorism* (the number of deaths in terrorist attacks worldwide between 1968-02 and 2006-06), *surnames* (frequencies of surnames in the 1990 US Census), *metabolic* (vertex degrees in the metabolic network of E. coli). We also include the *DBLPv12* dataset that consists of citation counts to over 4 million papers in the field of computer and information sciences sourced from (Tang et al., 2008)[3] as well as the citation data from the *RePEc*[4] database which features almost 2 million papers in economics (only papers cited at least once were included). Moreover, we consider the following information about users and questions on the Stack Overflow[5] site: the number of times each user profile and question was viewed, and the number of up- and down-votes each user has cast (only items greater than 0 were included, as new users cannot vote until they reach a certain reputation level).

Fig. 6 presents the empirical and fitted (based on the sample Gini indices) Lorenz curves. Let us note that all datasets' $N$s are of a considerable order of magnitude. Therefore, in all 15 cases, we can rely on the asymptotic formula (8) with little loss in precision. Our model is a very good fit (RMSE < 0.02) in six cases, but is unsatisfactory (RMSE > 0.045) in three instances.

### 4.2. Generalised Pareto distributions

Now the problem is to identify the probability distribution with CDF $F$ whose *Lorenz curve coincides with that given by* $L^{(\infty,G)}$, $0 < G < 1$. Let us recall (see Gastwirth, 1971) that:

$$L^{(\infty,G)}(u) = \frac{\int_0^u F^{-1}(t)\,dt}{\int_0^1 F^{-1}(t)\,dt} = \frac{\int_0^u F^{-1}(t)\,dt}{m}, \qquad (9)$$

where $m$ is the expected value, $m = \int_0^1 F^{-1}(t)\,dt = \int_0^\infty (1 - F(x))\,dx$.

The sought CDF $F$ can be obtained by inverting the quantile function $F^{-1}$ calculated by taking the derivative of $L^{(\infty,G)}(u)$ w.r.t. $u$. After some basic algebraic transformations, see Appendix A.5, we arrive at the following equation:

---

[2] See https://github.com/aaronclauset/aaronclauset.github.io/tree/master/powerlaws/data; current as of 2023-11-01. We skipped the *weblinks* dataset whose size $N$ is too large (hundreds of millions).

[3] See https://www.aminer.org/citation; database snapshot dated 2020-04-09.

[4] We thank Jose Manuel Barrueco for providing us with a large snapshot of RePEc data on 2022-11-03. All data are freely available for download at http://citec.repec.org/api.html.

[5] Stack Overflow, see https://stackoverflow.com/, is a very popular site where users build a knowledge base related to programming with 24M questions, 35M answers and 21M registered users. We use data dumps dated 2023-09-12 that we downloaded from https://archive.org/details/stackexchange.

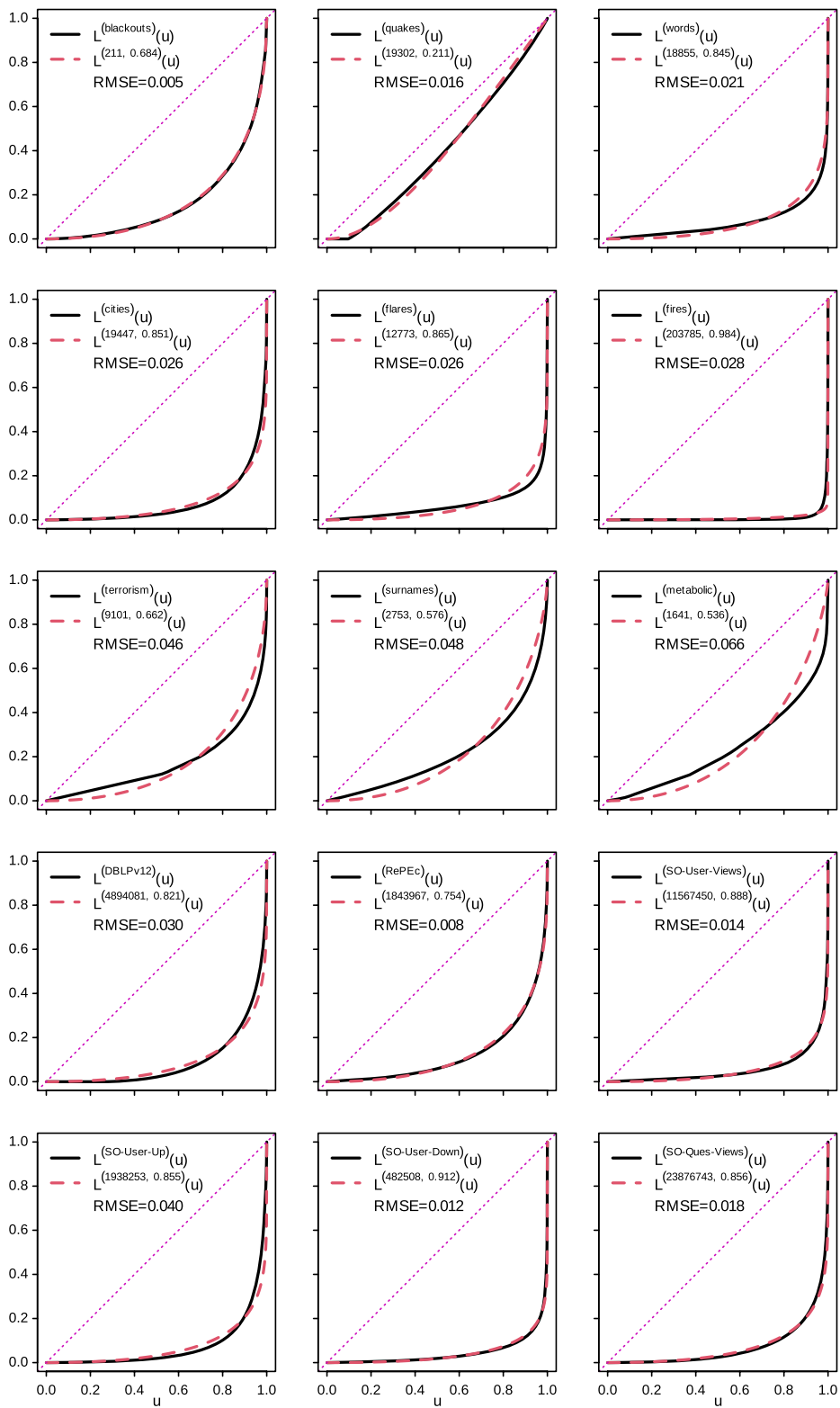

**Fig. 6.** The empirical and fitted curves for a wide range of datasets from different domains, including bibliometrics, informetrics, and environmental sciences; see Examples 4 and 6. Due to each *N*'s being large, we can replace $L^{(N,G)}$ with $L^{(\infty,G)}$ with little loss in precision.

$$F(x) = \begin{cases} \left(1 - \left(1 + \frac{x}{\sigma}\right)^{-\alpha}\right) I[0 \leqslant x \leqslant -\sigma] + I[x > -\sigma], & \text{if } 0 < G < \frac{1}{2}, \\ (1 - e^{-mx}) I[x > 0], & \text{if } G = \frac{1}{2}, \\ \left(1 - \left(1 + \frac{x}{\sigma}\right)^{-\alpha}\right) I[x > 0], & \text{if } \frac{1}{2} < G < 1. \end{cases} \quad (10)$$

where $\alpha = \frac{G}{2G-1}$ and $\sigma = m(\alpha - 1) = m\frac{1-G}{2G-1}$, and $I$ is the indicator function. Let us note that for $G < 1/2$ it holds $-\sigma > 0$. In this case, the support is finite.

The above corresponds to the CDF of the Pickands Generalised Pareto Distribution (GPD; Pickands, 1975). As it is well-known, this distribution family unifies three different models (see Hosking & Wallis, 1987, Arnold, 2015, p. 11, and Johnson et al., 1994, p. 614). Namely, depending on the value of the parameter $G$, here are the possible cases for a random variable $X$ with the CDF given by Eq. (10):

(i) Case $0 < G < \frac{1}{2}$ — *Scaled Beta Distribution*

Here, $\alpha < 1$ and so the support of the random variable becomes bounded. By setting $\tau = -\sigma = m\frac{1-G}{1-2G} > 0$ as the free scale parameter, we find the formula for the CDF:

$$F(x) = 1 - \left(1 - \frac{x}{\tau}\right)^{-\alpha}, \quad 0 < x < \tau, \tau > 0,$$

with the corresponding density function:

$$f(x) = \frac{\tau^{-1}(1 - x\tau^{-1})^{-\alpha-1}}{-1/\alpha} = \frac{\tau^{-1}(1 - x\tau^{-1})^{-\alpha-1}}{B(1, -\alpha)}, \quad 0 < x < \tau, \tau > 0,$$

where $B$ denotes the beta function. This can be viewed as a scaled Beta distribution of the first kind with shape parameters 1 and $\frac{G}{1-2G}$ (Marshall & Olkin, 2007, p. 494) or a uniform distribution with scale and frailty parameters (Marshall & Olkin, 2007, p. 672).

As a very special case, for $G = \frac{1}{3}$ ($\alpha = -1$), we obtain the CDF of the continuous uniform distribution on the interval $(0, \tau)$, i.e., $F(x) = \frac{x}{\tau}$, $0 < x < \tau$.

(ii) Case $G = \frac{1}{2}$ — *Exponential Distribution*

Here we find the CDF of the exponential distribution with mean equal to $m$:

$$F(x) = 1 - \exp\left(\frac{-x}{m}\right), x > 0.$$

(iii) Case $\frac{1}{2} < G < 1$ — *Pareto Type II Distribution*

Here, $\alpha > 1$ (which guarantees that the expected value is well-defined), $\sigma > 0$ and $F$ takes the form of the CDF of a Pareto Type II distribution with shape parameter $\alpha$ and scale parameter $\sigma > 0$. Such a random variable is usually denoted by $X \sim P(\text{II})(0, \sigma, \alpha)$ (e.g., Arnold, 2015, p. 151).

Note that under the Pareto Type II distribution, the Gini index cannot fall below the minimum level of $1/2$, which is consistent with our derivations (compare also Biró et al., 2023).

In conclusion, the family of Lorenz curves $L^{(\infty, G)}$, $0 < G < 1$, contains the Lorenz curves generated by all and only Pickands' Generalised Pareto Distributions for which the mean exists (as otherwise the Lorenz curve is not defined).

Note that these formulae generalise some of the cases studied in Balakrishnan et al. (2010). See also Siudem et al. (2022), where a different parametrisation of the Pareto Type II distribution was obtained in the limit.

**Example 5.** Continuing the countrywise family income study (Example 3), let us compare our new finite-sample model $L^{(N, G)}$ against its asymptotic version $L^{(\infty, G)}$ being an alternative parametrisation of GPD. Furthermore, let us consider the Lotkaian model (e.g., Egghe, 2005a), which in our context is equivalent to the power law (e.g., Clauset et al., 2009) and the classic Pareto (Type I) distribution (e.g., Arnold, 2015). Its Lorenz curve is given by $L_1^{(\alpha)}(u) = 1 - (1-u)^{1-1/\alpha}$, which we can also reparameterise based on the Gini index knowing that $G = 1/(2\alpha - 1)$, leading to the family of curves we denote by $L_1^{(G)}$.

Table 1 gives the root mean squared errors between the empirical and fitted Lorenz curves in the case where the estimation is solely based on the Gini index. Furthermore, in parentheses, we list the minimal possible RMSEs (where the model parameter is optimised for numerically so as to minimise this error metric).

First, we note that in each case, the finite-sample model $L^{(N, G)}$ is better than the remaining ones. The obtained RMSEs are very close to the optimal one. This indicates that the sample Gini index is a well-behaving estimator of the theoretical $G$ parameter. Second, we note that the $L^{(\infty, G)}$-based (GPD) model is better than the Lotkaian one, sometimes considerably. Also in this case, using the sample Gini index as an estimator of the $G$ parameter, gives quite satisfactory results.

**Table 1**

Comparison of three models of Lorenz curves parameterised by the Gini index (where it is taken from each sample) in terms of the root mean squared error for the countrywise family income data (Examples 3 and 5). In round brackets, minimum achievable RMSE for each model is given. The new finite-sample model outperforms the other ones. In this case, using the sample Gini index to estimate $G$ works almost equally well as minimising RMSE numerically.

|    | N  | G    | RMSE (Min RMSE) | | |
|----|----|------|-----------------|---|---|
|    |    |      | $L^{(N,G)}$ | $L^{(\infty,G)}$ | $L_1^{(G)}$ |
| AU | 10 | 0.38 | **0.013** (0.012) | 0.023 (0.018) | 0.055 (0.045) |
| CA | 10 | 0.38 | **0.012** (0.012) | 0.023 (0.018) | 0.055 (0.046) |
| NL | 10 | 0.34 | **0.022** (0.021) | 0.029 (0.027) | 0.044 (0.036) |
| NO | 10 | 0.34 | **0.012** (0.012) | 0.022 (0.018) | 0.053 (0.044) |
| SE | 10 | 0.32 | **0.022** (0.021) | 0.029 (0.028) | 0.042 (0.035) |
| CH | 10 | 0.42 | **0.027** (0.027) | 0.034 (0.031) | 0.044 (0.034) |
| DE | 10 | 0.34 | **0.021** (0.020) | 0.029 (0.027) | 0.044 (0.036) |
| UK | 10 | 0.38 | **0.013** (0.013) | 0.023 (0.018) | 0.054 (0.045) |
| US | 10 | 0.40 | **0.009** (0.009) | 0.022 (0.014) | 0.060 (0.049) |

**Table 2**

Comparison of the asymptotic (GPD) and Lotkaian (power-law) models on a wide variety of datasets from different domains (Examples 4 and 6).

|               | N          | G    | RMSE (Min RMSE) | |
|---------------|------------|------|------------------|---|
|               |            |      | $L^{(\infty,G)}$ | $L_1^{(G)}$ |
| blackouts     | 211        | 0.68 | **0.006** (0.006) | 0.061 (0.054) |
| quakes        | 19,302     | 0.21 | **0.016** (0.016) | 0.045 (0.042) |
| words         | 18,855     | 0.84 | **0.021** (0.015) | **0.021** (0.020) |
| cities        | 19,447     | 0.85 | **0.026** (0.023) | 0.060 (0.052) |
| flares        | 12,773     | 0.86 | 0.026 (0.021) | **0.023** (0.022) |
| fires         | 203,785    | 0.98 | **0.028** (0.025) | 0.031 (0.029) |
| terrorism     | 9,101      | 0.66 | 0.046 (0.035) | **0.021** (0.020) |
| surnames      | 2,753      | 0.58 | 0.048 (0.039) | **0.025** (0.023) |
| metabolic     | 1,641      | 0.54 | 0.066 (0.059) | **0.011** (0.010) |
| DBLPv12       | 4,894,081  | 0.82 | **0.030** (0.024) | 0.070 (0.061) |
| RePEc         | 1,843,967  | 0.75 | **0.008** (0.008) | 0.053 (0.047) |
| SO-User-Views | 11,567,450 | 0.89 | **0.014** (0.014) | 0.035 (0.031) |
| SO-User-Up    | 1,938,253  | 0.86 | **0.040** (0.035) | 0.072 (0.064) |
| SO-User-Down  | 482,508    | 0.91 | **0.012** (0.011) | 0.029 (0.026) |
| SO-Ques-Views | 23,876,743 | 0.86 | **0.018** (0.016) | 0.051 (0.044) |

**Example 6.** Let us go back to Example 4, where we have studied 15 datasets from various domains, including informetrics and environmental sciences. In each case, the sample size $N$ is large, hence $L^{(N,G)} \simeq L^{(\infty,G)}$. We shall thus only compare the asymptotic model $L^{(\infty,G)}$ with the Lotkaian one, $L_1^{(G)}$ (as given in Example 5). From Table 2 we see that where the $L^{(\infty,G)}$ model is a good fit, the RMSE in the case where $G$ is estimated by means of the sample Gini index is close to the optimal one. Moreover, in the three cases where our model does not fit data well (*terrorism, surnames, metabolic*), $L_1^{(G)}$ with $G$ taken from data is actually quite satisfactory.

## 5. Conclusion

In the context of a general information production process, which is a formal mechanism in which sources produce items (Egghe, 1990 and Egghe, 2005b, p. 8), the Gini-stable process can be viewed as a model where in each step, a new source is added under the special condition of keeping the concentration, as measured by the Gini index, unchanged. In this sense, the Gini-stable model derived herein can *also* be interpreted as a process that describes how production changes *over time* in the sense invoked by (Burrell, 1992).

It is interesting to note that a model mathematically equivalent to Eq. (5) for $G > 1/2$ (minus the scaling of the vectors' elements) was already featured in (Siudem et al., 2020) and in (Gagolewski et al., 2022); see Appendix A.2 for more details. Their parametrisation is defined via the value $\rho > 0$ explaining the preferential attachment mechanism, i.e., an author's specific tendency to produce articles more or less attractive to citations. Also notice that the recent paper by Biró et al. (2023) suggests that $G > 1/2$ is the most prevalent case for bibliometric data; compare Table 2.

In the literature, many parametric families of *continuous* Lorenz curves are known. In the present article, we derived an explicit expression for the Lorenz curve of a probability vector generated by the Gini-stable process defined by Eq. (5), in its genuine discrete formulation. Our formula can also be seen as a model for an *empirical* Lorenz curve, that is, for Lorenz curves of empirical distributions of *finite* samples. In this sense, the family presented here is a unique model in its genre. Within the context of a general information production process, such Lorenz curves represent the cumulative proportion of total production of all sources as ordinate against the cumulative proportion of sources as abscissa (in particular, one obtains the Lorenz curve of the productivity if the sources are arranged in increasing order of productivity). It is important to emphasise the fact that we refer here to the case of the size-frequency domain (Egghe & Waltman, 2011).

We proved that the Lorenz curve gives a complete ordering relationship among the probability vectors of the Gini-stable process, with respect to its parameters $N$ and $G$ – where the former represents the number of sources and the latter the value of the Gini index within its natural range $(0, 1)$. In particular, probability vectors of unequal lengths turn out to be comparable with respect to the Lorenz ordering for suitable values of parameter $G$.

We also derived and studied the asymptotic form of the Lorenz curve when $N$ goes to infinity, obtaining a new parametric family of Lorenz curves that is ordered with respect to the parameter $G$. Up to a change in scale, each Lorenz curve in this family characterises an absolutely continuous distribution on $(0, \infty)$, with finite mean, which we have shown to belong to the Pickands Generalised Pareto Distribution family with a new parametrisation that relies directly on $G$. This result sheds a new light on the genesis of this family of distributions. Importantly, the total ordering of our Lorenz curve family is a crucial property thereof because, in general, the Lorenz curves can cross (the Lorenz order is only a partial order).

Lastly, we showed our model fits well to a few informetric datasets. The advantage of our parametrisation is that it only relies on the sample Gini index, which is very easy to compute. No sophisticated curve fitting is necessary.

In a follow-up work (Bertoli-Barsotti et al., 2023), we derive the formulae for the Bonferroni, Hoover, and De Vergottini indices in our new model so that we can recreate their values based on the value of the Gini index. Future research will involve the study of the processes preserving other inequality, evenness, fairness, or spread measures; compare, e.g., (Beliakov et al., 2016, Bu, 2022, Chan, 2022, Gagolewski, 2015, Prathap, 2023, Prathap & Rousseau, 2023). It will be interesting to compare our finite-sample model with other parametric families of Lorenz curves, e.g., ones reviewed in (Chotikapanich, 2008).

**CRediT authorship contribution statement**

**Lucio Bertoli-Barsotti:** Writing – review & editing, Writing – original draft, Methodology, Conceptualization. **Marek Gagolewski:** Writing – review & editing, Writing – original draft, Visualization, Software, Methodology, Conceptualization. **Grzegorz Siudem:** Writing – review & editing, Writing – original draft, Methodology, Conceptualization. **Barbara Żogała-Siudem:** Writing – review & editing, Writing – original draft, Visualization, Software.

**Declaration of competing interest**

The authors certify that they have no affiliations with or involvement in any organisation or entity with any financial interest or non-financial interest in the subject matter or materials discussed in this manuscript.

**Acknowledgements**

This research was supported by the Australian Research Council Discovery Project ARC DP210100227 (MG). We thank the anonymous reviewers for their detailed remarks which helped improve the manuscript.

**Appendix A. Proofs**

*A.1. Derivation of the formula for $p_i^{(N,G)}$*

To show that the iterative formula (4) leads to Eq. (5), let us re-write recurrence definition of $p^{(N,G)}$ for $i \leqslant N$. Assuming for brevity that $p_i^{(N,G)} = 0$ if $i > N$, we get:

$$\begin{aligned}
p_i^{(N,G)} &= b_{N-1} p_i^{(N-1,G)} + a_{N-1} \\
&= b_{N-1} \left( b_{N-2} p_i^{(N-2,G)} + a_{N-2} \right) + a_{N-1} \\
&= b_{N-1} b_{N-2} p_i^{(N-2,G)} + a_{N-2} b_{N-1} + a_{N-1} \\
&= b_{N-1} b_{N-2} b_{N-3} p_i^{(N-3,G)} + a_{N-3} b_{N-1} b_{N-2} + a_{N-2} b_{N-1} + a_{N-1} = \ldots \text{etc.} \ldots \\
&= \left( \prod_{\ell=1}^{N-i+1} b_{N-\ell} \right) \underbrace{p_i^{(i-1,G)}}_{=0} + \sum_{k=1}^{N-i+1} a_{N-k} \prod_{\ell=1}^{k-1} b_{N-\ell} = \sum_{k=i-1}^{N-1} a_k \prod_{\ell=k+1}^{N-1} b_\ell \\
&\stackrel{\text{Eq. (4)}}{=} \sum_{k=i-1}^{N-1} \frac{1-G}{(k+1)(kG+1-G)} \prod_{\ell=k+1}^{N-1} \frac{\ell G}{\ell G + 1 - G} = \frac{1-G}{G} \sum_{k=i-1}^{N-1} \frac{1}{(k+1)(k+\frac{1-G}{G})} \prod_{\ell=k+1}^{N-1} \frac{\ell}{\ell + \frac{1-G}{G}} \\
&= \frac{1-G}{G} \sum_{k=i-1}^{N-1} \frac{\prod_{\ell=k+2}^{N-1} \ell}{\prod_{\ell=k}^{N-1} \left( \ell + \frac{1-G}{G} \right)} = \frac{1-G}{G} \sum_{k=i-1}^{N-1} \frac{\prod_{\ell=k+2}^{N-1} \ell}{\prod_{\ell=k}^{N-1} \left( \ell + \frac{1-G}{G} \right)}
\end{aligned}$$

$$
\begin{aligned}
&= \frac{1-G}{G} \sum_{k=i-1}^{N-1} \frac{\Gamma(N)/\Gamma(k+2)}{\Gamma\left(N + \frac{1-G}{G}\right)/\Gamma\left(k + \frac{1-G}{G}\right)} \\
&\stackrel{\star}{=} (1-\rho) \frac{\Gamma(N)}{\Gamma(N+1-\rho)} \sum_{k=i}^{N} \frac{\Gamma(k-\rho)}{\Gamma(k+1)} = \frac{1-\rho}{N} \frac{\Gamma(N+1)}{\Gamma(N+1-\rho)} \sum_{k=i}^{N} \frac{\Gamma(k-\rho)}{\Gamma(k+1)},
\end{aligned}
\quad (11)
$$

where in $\star$, we apply the substitution $G = \frac{1}{2-\rho}$ (compare Bertoli-Barsotti, 2023). Equation (11) is equivalent (with a slightly different notation) to Eq. (8) in (Siudem et al., 2020), where $p_i^{(N,G)}$ is given by Eq. (11) therein, with $C=1$ and $\rho = 2 - \frac{1}{G}$, which results exactly in our Eq. (5) for $G \neq \frac{1}{2}$ ($\rho \neq 0$). The case of $G = \frac{1}{2}$ ($\rho = 0$) in Eq. (5), which we can express by means of harmonic numbers, must have been treated separately to avoid the division by 0; see Eq. (8) in (Cena et al., 2022).

*A.2. Proof of Proposition 1*

It is sufficient to prove that for any $N$ and $G_1 < G_2$, it holds $\Sigma_k^{(N,G_1)} < \Sigma_k^{(N,G_2)}$ for all $k = 1, \ldots, N-1$. In our setting, this holds if and only if all the partial derivatives $\partial \Sigma_k^{(N,G)}/\partial G$ are nonnegative. We have:

$$
\frac{\partial \Sigma_k^{(N,G)}}{\partial G} = \frac{-G + (2G-1)\left(H_{N-2+\frac{1}{G}} - H_{k-2+\frac{1}{G}}\right)}{G(2G-1)^2} \frac{\Gamma(N)}{\Gamma(k)} \frac{\Gamma(k-1+1/G)}{\Gamma(N-1+1/G)} + \frac{k}{(2G-1)^2 N} = \underbrace{\frac{1}{(2G-1)^2}}_{>0} f_{k,N}(G),
$$

with

$$
f_{k,N}(G) = \frac{\Gamma(N)\Gamma(k-1+1/G)}{G\,\Gamma(k)\Gamma(N-1+1/G)} \left(-G + (2G-1)\left(H_{N-2+\frac{1}{G}} - H_{k-2+\frac{1}{G}}\right)\right) + \frac{k}{N}.
$$

We have $f_{k,N}\left(\frac{1}{2}\right) = -k/N + k/N = 0$ and for $G \neq \frac{1}{2}$ it holds:

$$
\frac{\partial f_{k,N}(G)}{\partial G} = \frac{(2G-1)\Gamma(N)\Gamma(k-1+1/G)}{G^3 \Gamma(k)\Gamma(N-1+1/G)}\left[\left(H_{k-2+\frac{1}{G}} - H_{N-2+\frac{1}{G}}\right)^2 + \psi'\left(k-1+\frac{1}{G}\right) - \psi'\left(N-1+\frac{1}{G}\right)\right],
$$

where $\psi(x) = \Gamma'(x)/\Gamma(x)$ is the polygamma function. Since the gamma function is non-negative for positive arguments, and the first derivative of the polygamma function $\psi'(x)$ is decreasing in $x$, the sign of $\partial f_{k,N}(G)/\partial G$ depends on the term $2G-1$. In particular:

$$
\frac{\partial f_{k,N}(G)}{\partial G} < 0 \text{ for } G < \frac{1}{2},
$$

$$
\frac{\partial f_{k,N}(G)}{\partial G} > 0 \text{ for } G > \frac{1}{2},
$$

which proves that $f_{k,N}(G) > 0$ for $G \neq \frac{1}{2}$ and $f_{k,N}\left(\frac{1}{2}\right) = 0$. However, since $1/(2G-1)^2$ is singular in $G = \frac{1}{2}$, we observe that:

$$
\frac{\partial \Sigma_k^{(N,1/2)}}{\partial G} = \frac{2\Gamma(k+1)\Gamma(N)}{\Gamma(k)\Gamma(N+1)}\left((H_k - H_N)^2 + \psi^{(1)}(k+1) - \psi^{(1)}(N+1)\right) > 0,
$$

for similar reasons to the ones discussed above. Thus, we have shown that $\partial \Sigma_k^{(N,G)}/\partial G > 0$ for every $G$. One gets $\Sigma_k^{(N,G_1)} < \Sigma_k^{(N,G_2)}$ for $k < N$, and obviously $\Sigma_N^{(N,G_1)} = \Sigma_N^{(N,G_2)} = 1$, which proves our proposition for $u = \frac{i}{N}$ for $i = 1, \ldots, N-1$. Since the Lorenz curve is defined as a linear interpolation between these points, the ordering of interest is preserved everywhere, and the proof is complete.

*A.3. Proof of Proposition 2*

For brevity of notation, let us recall the relationship between the Leimkuhler and the Lorenz curves: $K(u) = 1 - L(1-u)$, $0 \leq u \leq 1$. In our case, we find:

$$
K^{(N,G)}\left(\frac{i}{N}\right) = 1 - L^{(N,G)}\left(1 - \frac{i}{N}\right) = \Sigma_i^{(N,G)}, \ i = 0, 1, \ldots, N.
$$

The *Leimkuhler order* is $p \leq_K q$ if and only if $K^p(u) \leq K^q(u)$ for all $0 \leq u \leq 1$. As the Leimkuhler order is equivalent to the Lorenz order, the thesis of Proposition 2:

$$
L^{(N_1,G)}(u) \geq L^{(N_2,G)}(u), \ \text{ for every } 0 \leq u \leq 1, \ N_1 < N_2,
$$

is equivalent to:

$$K^{(N_1,G)}(u) \leq K^{(N_2,G)}(u), \text{ for every } 0 \leq u \leq 1, N_1 < N_2.$$

By construction, the Leimkuhler curve $K^{(N,G)}$ is defined by straight-line segments with slopes given by the ratio between the component of the vector $p^{(N,G)}$ and the mean, that is, for $i = 0, 1, \ldots, N$:

$$\frac{p_i^{(N,G)}}{\frac{1}{N}} = N p_i^{(N,G)} = \begin{cases} \frac{1-G}{2G-1}\left(\frac{\Gamma(N+1)\Gamma(i-2+1/G)}{\Gamma(N+1/G-1)\Gamma(i)} - 1\right), & \text{for } G \neq \frac{1}{2}, \\ H_N - H_{i-1}, & \text{for } G = \frac{1}{2}. \end{cases}$$

To better illustrate the behaviour of the Leimkuhler curve as the number of the line segments grows from $N$ to $(N + 1)$, we will prove the following result about their slopes.

(a) We will show that the following inequalities hold:

$$N p_i^{(N,G)} < (N+1) p_i^{(N+1,G)}, \qquad i = 1, 2, \ldots, N, \tag{12}$$

$$(N+1) p_i^{(N+1,G)} < N p_{i-1}^{(N,G)}, \qquad i = 2, \ldots, N+1. \tag{13}$$

For $G = 1/2$, the inequality (12) is equivalent to $(H_N - H_{i-1}) < (H_{N+1} - H_{i-1})$ which is clearly true.
On the other hand, for $G \neq 1/2$, the inequality (12) is equivalent to:

$$\frac{c-1}{2-c}\left(\frac{\Gamma(N+2)\Gamma(i-2+c)}{\Gamma(N+c)\Gamma(i)} - 1\right) > \frac{c-1}{2-c}\left(\frac{\Gamma(N+1)\Gamma(i-2+c)}{\Gamma(N+c-1)\Gamma(i)} - 1\right),$$

where $c = 1/G > 1$. There are two cases.
If $\frac{c-1}{2-c} > 0$, that is $c > 2$ (or $G > \frac{1}{2}$), the above inequality is equivalent to:

$$\frac{\Gamma(N+2)\Gamma(i-2+c)}{\Gamma(N+c)\Gamma(i)} > \frac{\Gamma(N+1)\Gamma(i-2+c)}{\Gamma(N+c-1)\Gamma(i)},$$

which is in turn equivalent to:

$$\frac{(N+1)\Gamma(N+1)}{(N+c-1)\Gamma(N+c-1)\Gamma(i)} > \frac{\Gamma(N+1)}{\Gamma(N+c-1)\Gamma(i)},$$

that is, $N + 1 > N + c - 1$, that holds true under the above condition on the parameter $c$.
If $\frac{c-1}{2-c} < 0$, that is $c > 2$ (or $G < \frac{1}{2}$), the inequality to be proved holds if and only if:

$$\left(\frac{\Gamma(N+2)\Gamma(i-2+c)}{\Gamma(N+c)\Gamma(i)} - 1\right) < \left(\frac{\Gamma(N+1)\Gamma(i-2+c)}{\Gamma(N+c-1)\Gamma(i)} - 1\right),$$

that easily leads to $N + c - 1 > N + 1$, that holds true under the condition $c > 2$.

Let us consider now the second inequality, (13). For $G = 1/2$, it is equivalent to $(H_{N+1} - H_{i-1}) < (H_N - H_{i-2})$. By definition of $H_i$, we rewrite it as $\sum_{j=i}^{N+1} 1/j < \sum_{j=i-1}^{N} 1/j$ iff $1/(N+1) < 1/(i-1)$, therefore it is true.
For $G \neq 1/2$, the inequality (13) holds if and only if:

$$\frac{c-1}{2-c}\left(\frac{\Gamma(N+2)\Gamma(i-2+c)}{\Gamma(N+c)\Gamma(i)} - 1\right) < \frac{c-1}{2-c}\left(\frac{\Gamma(N+1)\Gamma(i-3+c)}{\Gamma(N+c-1)\Gamma(i-1)} - 1\right), \qquad i = 2, \ldots, N+1.$$

If $\frac{c-1}{2-c} > 0$, that is $c < 2$ (or $G > \frac{1}{2}$), the above inequality can be simplified to:

$$\frac{\Gamma(N+2)\Gamma(i-2+c)}{\Gamma(N+c)\Gamma(i)} < \frac{\Gamma(N+1)\Gamma(i-3+c)}{\Gamma(N+c-1)\Gamma(i-1)},$$

which is in turn equivalent to:

$$\frac{(N+1)(i-3+c)}{(N+c-1)(i-1)} < 1,$$

that holds true if $c < 2$.
For $\frac{c-1}{2-c} < 0$ ($G < \frac{1}{2}$), the reasoning is similar to the one above.

(b) We will show that every vertex $\left(\frac{i}{N}, K^{(N,G)}\left(\frac{i}{N}\right)\right)$, $i = 1, \ldots, N-1$, of $K^{(N,G)}$ belongs to $K^{(N+1,G)}$.
More precisely, the thesis (b) is that:

$$K^{(N,G)}\left(\frac{i}{N}\right) = K^{(N+1,G)}\left(\frac{i}{N+1}\right) + \left(\frac{i}{N} - \frac{i}{N+1}\right)\frac{p_{i+1}^{(N+1,G)}}{(N-1)^{-1}}, \qquad i = 1, \ldots, N-1,$$

which is equivalent to:

$$K^{(N,G)}\left(\frac{i}{N}\right) = K^{(N+1,G)}\left(\frac{i}{N+1}\right) + \frac{i}{N} p_{i+1}^{(N+1,G)}, i = 1, \ldots, N-1.$$

The latter equation can be replaced by:

$$\frac{N}{i}\left(\Sigma_i^{(N,G)} - \Sigma_i^{(N+1,G)}\right) = p_{i+1}^{(N+1,G)}, \qquad i = 1, \ldots, N-1.$$

The required result follows after simple manipulations. Indeed, for $G = 1/2$, we find:

$$\frac{N}{i}\left(\Sigma_i^{(N,G)} - \Sigma_i^{(N+1,G)}\right) = \frac{N}{i}\left(\frac{i}{N}(1 + H_N - H_i) - \frac{i}{N+1}(1 + H_{N+1} - H_i)\right)$$

$$= \frac{1}{N+1} + H_N - H_i - \frac{N}{N+1}(H_{N+1} - H_i)$$

$$= \frac{1}{N+1}\left(H_{N+1} - H_i\right).$$

Moreover, for $G \neq 1/2$, we get:

$$\frac{N}{i}\left(\Sigma_i^{(N,G)} - \Sigma_i^{(N+1,G)}\right) = \frac{N}{i}\frac{G}{(2G-1)}\frac{i}{N}\left(\frac{\Gamma(N+1)\Gamma(i-1+1/G)}{\Gamma(i+1)\Gamma(N-1+1/G)} + 1 - 1/G\right)$$

$$-\frac{N}{i}\frac{G}{(2G-1)}\frac{i}{(N+1)}\left(\frac{\Gamma(N+2)\Gamma(i-1+1/G)}{\Gamma(i+1)\Gamma(N+1/G)} + 1 - 1/G\right)$$

$$= \frac{(G-1)}{(2G-1)(N+1)} + \frac{G\Gamma(N+1)\Gamma(i-1+1/G)}{(2G-1)\Gamma(i+1)\Gamma(N-1+1/G)}\left(1 - \frac{N}{N-1+1/G}\right)$$

$$= -\frac{(1-G)}{(2G-1)(N+1)} + \frac{(1-G)\Gamma(N+1)\Gamma(i-1+1/G)}{(2G-1)\Gamma(i+1)\Gamma(N+1/G)}$$

$$= \frac{(1-G)}{(2G-1)}\frac{1}{(N+1)}\left(\frac{\Gamma(N+2)\Gamma(i-1+1/G)}{\Gamma(N+1/G)\Gamma(i+1)} - 1\right),$$

where the last terms in both equations are equal to $p_{i+1}^{(N+1,G)}$.

We prove the thesis of the Proposition 2 by contradiction. Assume that the Leimkuhler curves $K^{(N,G)}$ and $K^{(N+1,G)}$ intersect. Then, there would be at least one vertex of $K^{(N,G)}$ not belonging to $K^{(N+1,G)}$, but this contradicts (b).

### A.4. Derivation of the formula for $L^{(\infty,G)}$

For any fixed $u$ such that $i = uN$, $i = 1, 2, \ldots, N$, we can express the Leimkuhler curve as:

$$K^{(N,\overline{\rho})}(u) = \frac{u}{1-\overline{\rho}}\left(\frac{\Gamma(Nu+\overline{\rho})\Gamma(N)}{u\Gamma(Nu)\Gamma(N+\overline{\rho})} - \overline{\rho}\right), \qquad 0 \leq u \leq 1, \overline{\rho} > 0, \overline{\rho} \neq 1,$$

where, to simplify the notation, we substitute $\overline{\rho} = 1/G - 1$. Since $\lim_{z \to \infty} \frac{\Gamma(z+\alpha)}{\Gamma(z)z^\alpha} = 1$ (Laforgia & Natalini, 2012), we find:

$$K^{(\infty,\overline{\rho})}(u) = \lim_{N \to \infty} K^{(N,\overline{\rho})}(u) = \frac{u}{1-\overline{\rho}}\left(u^{\overline{\rho}-1} - \overline{\rho}\right),$$

or, equivalently:

$$K^{(\infty,G)}(u) = \frac{G}{2G-1}u^{(1/G-1)} - \frac{1-G}{2G-1}u, \qquad 0 \leq u \leq 1, \ 0 < G < 1, \ G \neq \frac{1}{2}.$$

The case for $G = 1/2$ is obtained by taking the limit, using L'Hôpital's rule,

$$\lim_{G \to 1/2} K^{(\infty,G)}(u) = K^{(\infty,1/2)}(u) = u - u\log u.$$

Then, as $L^{(\infty,G)}(u) = 1 - K^{(\infty,G)}(1-u)$, after some simple transformations, we can arrive at Eq. (8).

### A.5. Derivation of the CDF F

To obtain the CDF $F$ given by Eq. (10), we start with the derivation of the formula for $\partial L^{(\infty,G)}(u)/\partial u$. Taking the derivative of the function given by Eq. (8), we get:

$$\frac{\partial L^{(\infty,G)}}{\partial u} = \begin{cases} 1 + \frac{G}{2G-1}\left(\frac{1-G}{G}(1-u)^{\frac{1}{G}-2} - 1\right), & \text{if } G \neq \frac{1}{2}, \\ -\log(1-u), & \text{if } G = \frac{1}{2}. \end{cases}$$

From Eq. (9), we see that $F(x)$ is given simply by $F^{-1}(x) = m\frac{\partial L^{(\infty,G)}}{\partial u}$, which, after some simple transformations, yields:

$$F(x) = \begin{cases} 1 - \left(\dfrac{(1-G)m}{2Gx - Gm - xm + m}\right)^{\frac{G}{2G-1}}, & \text{if } G \neq \frac{1}{2}, \\ 1 - e^{-mx}, & \text{if } G = \frac{1}{2}. \end{cases}$$

$$= \begin{cases} 1 - \left(\dfrac{2Gm^{-1}x - G - m^{-1}x + 1}{1 - G}\right)^{\frac{-G}{2G-1}}, & \text{if } G \neq \frac{1}{2}, \\ 1 - e^{-mx}, & \text{if } G = \frac{1}{2}. \end{cases}$$

$$= \begin{cases} 1 - \left(1 + \dfrac{(1-2G)}{G-1}\dfrac{x}{m}\right)^{\frac{-G}{2G-1}}, & \text{if } G \neq \frac{1}{2}, \\ 1 - e^{-mx}, & \text{if } G = \frac{1}{2}. \end{cases}$$

for the admissible range of $x$.

## References

Aaberge, R. (2001). Axiomatic characterization of the Gini coefficient and Lorenz curve orderings. *Journal of Economic Theory*, *101*, 115–132. https://doi.org/10.1006/jeth.2000.2749.

Argyris, N., Karsu, Ö., & Yavuz, M. (2022). Fair resource allocation: Using welfare-based dominance constraints. *European Journal of Operational Research*, *297*, 560–578. https://doi.org/10.1016/j.ejor.2021.05.003.

Arnold, B. C. (1991). Preservation and attenuation of inequality as measured by the Lorenz order. In *Stochastic orders and decision under risk* (pp. 25–37).

Arnold, B. C. (2015). *Pareto distributions* (2nd ed.). New York, NY, USA: Chapman and Hall/CRC.

Arnold, B. C., & Sarabia, J. M. (2018). *Majorization and the Lorenz order with applications in applied mathematics and economics*. Springer.

Balakrishnan, N., Sarabia, J. M., & Kolev, N. (2010). A simple relation between the leimkuhler curve and the mean residual life. *Journal of Informetrics*, *4*, 602–607.

Beliakov, G., Gagolewski, M., & James, S. (2016). Penalty-based and other representations of economic inequality. *International Journal of Uncertainty, Fuzziness and Knowledge-Based Systems*, *24*(Suppl. 1), 1–23. https://doi.org/10.1142/S0218488516400018.

Bertoli-Barsotti, L. (2023). Equivalent Gini coefficient, not shape parameter! *Scientometrics*, *128*, 867–870.

Bertoli-Barsotti, L., Gagolewski, M., Siudem, G., & Żogała Siudem, B. (2023). Equivalence of inequality indices: Three dimensions of impact revisited, https://doi.org/10.48550/arXiv.2304.07479. Under review (preprint).

Bertoli-Barsotti, L., & Lando, T. (2019). How mean rank and mean size may determine the generalised Lorenz curve: With application to citation analysis. *Journal of Informetrics*, *13*, 387–396.

Biró, T. S., Telcs, A., Józsa, M., & Néda, Z. (2023). Gintropic scaling of scientometric indexes. *Physica A. Statistical Mechanics and its Applications*, *618*, Article 128717. https://doi.org/10.1016/j.physa.2023.128717.

Bishop, J. A., Formby, J. P., & Smith, W. J. (1991). International comparisons of income inequality: Tests for Lorenz dominance across nine countries. *Economica*, *58*, 461–477.

Bu, N. (2022). A new fairness notion in the assignment of indivisible resources. *Mathematical Social Sciences*, *120*, 1–7. https://doi.org/10.1016/j.mathsocsci.2022.08.002.

Burrell, Q. L. (1992). The Gini index and the Leimkuhler curve for bibliometric processes. *Information Processing & Management*, *28*, 19–33.

Burrell, Q. L. (2005). Symmetry and other transformation features of Lorenz/Leimkuhler representations of informetric data. *Information Processing & Management*, *41*, 1317–1329.

Burrell, Q. L. (2006). On Egghe's version of continuous concentration theory. *Journal of the American Society for Information Science and Technology*, *57*, 1406–1411.

Burrell, Q. L. (2007). Egghe's construction of Lorenz curves resolved. *Journal of the American Society for Information Science and Technology*, *58*, 2157–2159.

Cena, A., Gagolewski, M., Siudem, G., & Żogała Siudem, B. (2022). Validating citation models by proxy indices. *Journal of Informetrics*, *16*, Article 101267. https://doi.org/10.1016/j.joi.2022.101267.

Chan, T. (2022). On a new class of continuous indices of inequality. *Mathematical Social Sciences*, *120*, 8–23. https://doi.org/10.1016/j.mathsocsci.2022.08.003.

Chotikapanich, D. (Ed.). (2008). *Modeling income distributions and Lorenz curves. Economic studies in equality, social exclusion and well-being: Vol. 5*. Springer.

Clauset, A., Shalizi, C., & Newman, M. (2009). Power-law distributions in empirical data. *SIAM Review*, *51*, 661–703.

Egghe, L. (1990). The duality of informetric systems with applications to the empirical laws. *Journal of Information Science*, *16*, 17–27.

Egghe, L. (2005a). *Journal of the American Society for Information Science and Technology*, *56*. https://doi.org/10.1002/asi.20157.

Egghe, L. (2005b). *Power laws in the information production process: Lotkaian informetrics*. Elsevier.

Egghe, L. (2005c). Zipfian and Lotkaian continuous concentration theory. *Journal of the American Society for Information Science and Technology*, *56*, 935–945.

Egghe, L., & Rousseau, R. (2023). Global informetric impact: A description and definition using bundles. *Journal of Informetrics*, *17*, Article 101366. https://doi.org/10.1016/j.joi.2022.101366.

Egghe, L., & Waltman, L. (2011). Relations between the shape of a size-frequency distribution and the shape of a rank-frequency distribution. *Information Processing & Management*, *47*, 238–245.

Gagolewski, M. (2015). Spread measures and their relation to aggregation functions. *European Journal of Operational Research*, *241*, 469–477. https://doi.org/10.1016/j.ejor.2014.08.034.

Gagolewski, M., Żogała Siudem, B., Siudem, G., & Cena, A. (2022). Ockham's index of citation impact. *Scientometrics*, *127*, 2829–2845. https://doi.org/10.1007/s11192-022-04345-2.

Gajdos, T. (2004). Single crossing Lorenz curves and inequality comparisons. *Mathematical Social Sciences*, *47*, 21–36.

Gastwirth, J. L. (1971). A general definition of the Lorenz curve. *Econometrica. Journal of the Econometric Society*, 1037–1039.

Gini, C. (1912). *Variabilità e mutabilità*. Bologna: C. Cuppini.

Hickey, R. J. (1983). Majorisation, randomness and some discrete distributions. *Journal of Applied Probability*, *20*, 897–902.

Hosking, J. R., & Wallis, J. R. (1987). Parameter and quantile estimation for the generalized Pareto distribution. *Technometrics*, *29*, 339–349.

Johnson, N. L., Kotz, S., & Balakrishnan, N. (1994). *Continuous multivariate distributions, volume 1: Models and applications*. John Wiley & Sons.

Laforgia, A., & Natalini, P. (2012). On the asymptotic expansion of a ratio of gamma functions. *Journal of Mathematical Analysis and Applications*, *389*, 833–837.

Magdalou, B. (2021). A model of social welfare improving transfers. *Journal of Economic Theory*, *196*, Article 105318. https://doi.org/10.1016/j.jet.2021.105318.

Marshall, A. W., & Olkin, I. (2007). *Life distributions: Structure of nonparametric, semiparametric, and parametric families*. Springer.

Marshall, A. W., Olkin, I., & Arnold, B. C. (2011). *Inequalities: Theory of majorization and its applications* (2nd ed.). Springer.

Pickands III, J. (1975). Statistical inference using extreme order statistics. *The Annals of Statistics*, *3*, 119–131.

Prathap, G. (2023). Inequality indices and their thermodynamic sensibilities, https://doi.org/10.13140/RG.2.2.19184.74246.

Prathap, G., & Rousseau, R. (2023). The modified repeat rate described within a thermodynamic framework. *Scientometrics, 128*, 3185–3195. https://doi.org/10.1007/s11192-023-04698-2.

Pyatt, G. (1976). On the interpretation and disaggregation of Gini coefficients. *The Economic Journal, 86*, 243–255.

Rousseau, R., Zhang, L., & Sivertsen, G. (2023). Using the weighted Lorenz curve to represent balance in collaborations: The BIC indicator. *Scientometrics, 128*, 609–622. https://doi.org/10.1007/s11192-022-04533-0.

Sarabia, J. M. (2008). A general definition of the Leimkuhler curve. *Journal of Informetrics, 2*, 156–163.

Siudem, G., Nowak, P., & Gagolewski, M. (2022). Power laws, the price model, and the Pareto type-2 distribution. *Physica A. Statistical Mechanics and its Applications, 606*, Article 128059. https://doi.org/10.1016/j.physa.2022.128059.

Siudem, G., Żogała-Siudem, B., Cena, A., & Gagolewski, M. (2020). Three dimensions of scientific impact. *Proceedings of the National Academy of Sciences, 117*, 13896–13900. https://doi.org/10.1073/pnas.2001064117.

Tang, J., et al. (2008). ArnetMiner: Extraction and mining of academic social networks. In *Proceedings of the fourteenth ACM SIGKDD international conference on knowledge discovery and data mining (SIGKDD'2008)* (pp. 990–998).

Tumen, S. (2011). Measuring earnings inequality: An economic analysis of the Bonferroni index. *The Review of Income and Wealth, 57*, 727–744.